\documentclass[letterpaper, 10 pt, conference]{ieeeconf}  

\IEEEoverridecommandlockouts                              
\usepackage{graphicx}
  \usepackage{here}
  \usepackage{array}
  \usepackage{color,amsmath,here}
  \usepackage{amssymb}
  \usepackage{enumerate}
  \usepackage{ascmac}
  \usepackage{tascmac}
  \usepackage{fancybox}
  \usepackage{float}
  \usepackage{amsmath}
  \usepackage{mathrsfs}

\newtheorem{proposition}{Proposition}

\newcounter{MYtempeqncnt}

\definecolor{mygreen}{rgb}{0, 0.7, 0}
\definecolor{myyellow}{rgb}{0.7, 0.7, 0}
\definecolor{mypurple}{rgb}{0.42, 0, 0.84}

\newcommand{\req}[1]{(\ref{#1})}

\title{\LARGE \bf
A Retrofitting-based Supplementary Controller Design for Enhancing
Damping Performance of Wind Power Systems
}

\author{Tomonori Sadamoto$^{1}$, Aranya Chakrabortty$^{2}$, Takayuki
Ishizaki$^{1}$, and Jun-ichi Imura$^{1}$
\thanks{
$^{1}$Department of Mechanical and Environmental Informatics, Graduate School of Information Science and Engineering, Tokyo Institute of Technology; 2-12-1, Meguro, Tokyo, Japan:}%
\thanks{$^2$Electrical \& Computer Engineering, North Carolina State
University; Raleigh, NC 27695}
\thanks{
\hspace{-3mm}{\tt\footnotesize \{sadamoto@cyb., ishizaki@,imura@\}mei.titech.ac.jp}
\hspace{-3mm}{\tt\footnotesize aranya.chakrabortty@ncsu.edu}
}%
}

\begin{document}

\maketitle
\thispagestyle{empty}
\pagestyle{empty}

\begin{abstract}
In this paper we address the growing concerns of wind power integration from the perspective of power system dynamics and stability. 
We propose a new retrofit control technique where an additional
 controller is designed at the doubly-fed induction generator site
 inside the wind power plant. This controller cancels the adverse
 impacts of the power flow from the wind side to the grid side on the
 dynamics of the overall power system. The main advantage of this controller is that it can be implemented by feeding back only the wind states and wind bus voltage without depending on any of the other synchronous machines in the rest of the system. Through simulations of a 4-machine Kundur power system model we show that the retrofit can efficiently enhance the damping performance of the system variable despite very high values of wind penetration.
\end{abstract}

\vspace{1mm}


\section{Introduction}\label{sec:intro}
With the rapid increase in wind penetration, power system operators
are gradually inclining towards developing controllers to mitigate
adverse impacts of doubly-fed induction generator (DFIG) models on
transient stability and small-signal stability of the grid \cite{gautam2009impact},\cite{chandra2015exploring}. The
current state-of-art method to mitigate such instability scenarios is by
heuristic tuning of PID controllers that are used for setpoint
regulation of the active and reactive power outputs of a DFIG. However,
this type of ad-hoc tuning may end up destabilizing the overall power
system due to incorrect choice of the PID gains, which will be shown
shortly in one of our simulation results. 
Thus, what operators need is a systematic control mechanism by which a state of the wind power plant can be regulated in a desired way without causing any grid instability,
and that too by preferably feeding back only the wind power plant state and wind bus voltage instead of relying on any of the other synchronous machines in the rest of the system.

In this paper, we fulfill this objective by designing a so-called {\it
retrofit} controller for a wind power plant, which is based on local
feedback of the wind power plant state only. 
The design methodology has been proposed in \cite{sadamoto2016retrofitting}. 
The advantages of the retrofit control are twofolds:
the retrofit controller is capable of enhancing the dynamic performance of the overall power system, and the controller design can be performed without explicit consideration of the dynamics of overall power system model. 
The design synthesis of the retrofit controller consists of three subsequent steps. First, we consider the wind-integrated system
without any retrofit controller, and ensure that the PSS gains of the
synchronous generators are robust enough to stabilize the overall power system
model for a given penetration level of wind power. Note that the
objective of this step is only to come up with a set of PSS gains that
guarantee stability; we do not require any PSS tuning here to optimize
the dynamic performance of the grid. Second, we consider the wind power plant to be isolated from the rest of the system, and connected to an
infinite bus, and design the retrofit controller using a linear
quadratic regulator (LQR) that depends on partial feedback of the wind
power plant state only. The controller is actuated through the current control
loop of the wind power plant in parallel to pre-existing PI controllers for setpoint
regulation of the currents. Finally, the synchronous generators with the
chosen PSS gains are integrated with the wind power plant with the
chosen LQR-based retrofit controller, and the combined grid dynamics are
shown to improve significantly. We validate our results by simulating
the classical 4-machine, 9-bus, 2-area Kundur power system model
\cite{kundur1994power} with a wind power plant at the intermediate bus.

The rest of this paper is organized as follows. In, Section \ref{sec:model}, 
we describe the coupled dynamic model of a power system with synchronous generators and a wind power plant, and establish the interaction between the two through their power flow equilibrium.
In Section \ref{sec:motiv}, we show a power system example that illustrates how a larger level of wind
penetration can induce oscillatory behavior in the line flows, and that incorrect PID tuning to counteract that oscillatory behavior may eventually end up destabilizing the system. In order to enhance the damping performance we propose a retrofit controller design in Section~\ref{sec:retrofit} based on the design method proposed in our recent paper \cite{sadamoto2016retrofitting}.
In Section~\ref{sec:numerc}, we investigate the efficiency of the retrofit control through a Kundur's 4-machine power system example with a wind power plant. Finally, concluding remarks are provided in Section~\ref{sec:concl}.

\vspace{1mm}

{\it Notation}:
We denote the imaginary unit by $j := \sqrt{-1}$, 
the set of real numbers by $\mathbb{R}$, the set of complex
numbers by $\mathbb{C}$.
The complex variables are described in bold fonts, e.g., ${\bf x}$. 
The conjugate of ${\bf x}$ is denoted by ${\bf x}^*$, and the absolute value of ${\bf x}$ by $|{\bf x}|$. 
A map $\mathcal{F}(\cdot)$ is said to be a dynamical map if the triplet $(x,u,y)$ with $y=\mathcal{F}(u)$ solves a system of differential equations
\[
\dot{x}=f(x,u),\quad y=g(x,u)
\]
with some functions $f(\cdot,\cdot)$ and $g(\cdot,\cdot)$, and an initial value $x(0)$.
In addition, the dynamical system $\mathcal{F}(\cdot)$ is said to be
stable if $\mathcal{F}(\cdot)$ is input-to-state stable \cite{khalil1996nonlinear}.
By the abuse of the terminology, the autonomous system $\dot{x} = f(x)$ is said to be stable if the system is asymptotically stable.
We denote the $\mathcal{L}_{2}$-norm of a square-integrable function $f(\cdot)$ by 
\[
\|f(t)\|_{\mathcal{L}_{2}}:=\sqrt{\textstyle{\int_{0}^{\infty}\|f(t)\|^{2}dt}}.
\]

\section{Motivating Example}\label{sec:motivating}
\subsection{Wind-integrated power system model}\label{sec:model}
\begin{figure}[t]
  \begin{center}
    \includegraphics[clip,width=85mm]{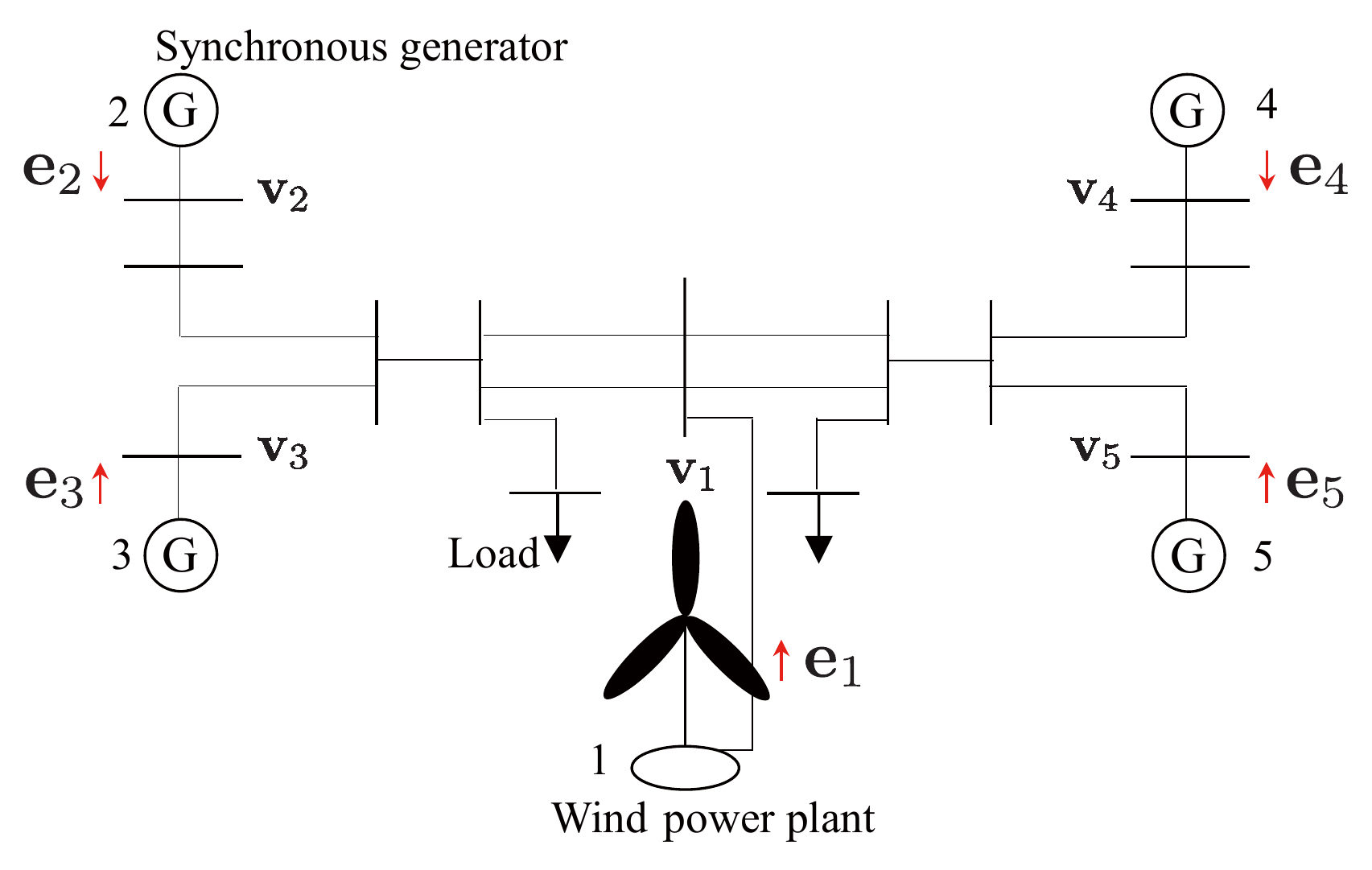}
    \caption{Wind-integrated power system where ${\bf v}_k$ is bus voltage and ${\bf e}_k$ is power outflowing from bus}
    \label{fig:example}
  \end{center}
\vspace{-5mm}
\end{figure}

We first develop a state-space model representing the electro-mechanical dynamics of a wind-integrated power system. While our modeling approach applies to any generic power system model, for the sake of illustration we refer to the 4-machine Kundur power system \cite{kundur1994power} with a wind plant at an intermediate bus as a running example. This model is shown in Fig. \ref{fig:example}. We refer to the total number of generation units, including both synchronous generators and a wind power plant by the symbol $N$. For the Kundur model, $N=5$.

Each generating unit generates active and reactive power depending on its bus
voltage, and injects the power to a grid satisfying power balance among
generation units and loads. 
The wind power plant model and synchronous generator
model are described in Section \ref{sec:wind_power} and
Section~\ref{sec:sync}, respectively, and the interconnection through
the grid is described in Section~\ref{sec:interconnection}.

\subsubsection{Wind Power Plants}\label{sec:wind_power}

We consider the wind power plant to be an aggregation of all the wind turbines and DFIGs inside it. Without loss of generality, we refer to the wind generator as generator 1 among the $N$ total generators in the system.  Following standard literature such as in \cite{akhmatov2002aggregate}, the model of all the generators are assumed to be identical for simplicity (the model can be easily extended to when this is not the case), and the total power injected into the grid from the wind plant is obtained by summing the power output of the individual generators.

The individual wind generator model has a wind turbine, DFIG, and an internal controller. 
The wind turbine is modeled as a two-inertia system of a rotor and generator \cite{sloth2010active}
\begin{equation}\label{wind_rotor}
 \dot{\eta} = A_{\eta}\eta + B_{\eta}\tau(i) + R_{\eta}(\omega_r)p_a,
  \quad \eta := [\omega_r, \omega_g, \theta]^{\sf T}
\end{equation}
where 
\[\hspace{-2mm}
\begin{array}{l}
  A_{\eta} = \left[
\begin{array}{ccc}
 -\frac{d_c+ d_r}{m_r} & \frac{d_c}{m_rn_g} & - \frac{k_c}{m_r}\\
 \frac{d_c}{m_gn_g} & -\frac{1}{m_g}\left(\frac{d_c}{n_g^2}+ d_g\right)&
  \frac{k_c}{m_gn_g}\\
 \bar{\omega}/2 & -\bar{\omega}/(2n_g) & 0
\end{array}
\right]\vspace{1mm}\\
 B_{\eta} = [0 ~ -2/(\bar{\omega}m_g)~0]^{\sf T}, ~
 R_{\eta}(\omega_r) = [2/(\bar{\omega}\omega_r m_r)~0~0]^{\sf T}, \quad
\end{array}
\]
$\omega_r$ and $\omega_g$ are the angular velocity of the rotor and generator,  
$\theta$ [rad] is the generator torsion angle, $\tau(i)$ is the torque
generated by the DFIG currents denoted by $i\in\mathbb R^4$, and
$p_a$ is the aerodynamic power input depending on wind speed. 
We assume $p_a$ to be constant since the time-scales over which wind speeds change by notable amounts are much slower than the time-scale of transient stability of the power system \cite{eping2005impact}, which is the main objective of interest for our controller design.
The positive constants, 
$m_r$,
$m_g$,
$d_r$, and
$d_g$,
are inertia and damping coefficients of the rotor and generator, 
$k_c$ and
$d_c$ are torsional stiffness and damping coefficients, 
$n_g$ is a gear ratio, and $\bar{\omega} = 120\pi$ [rad/s] is the synchronous frequency. 
The values of these various model parameters are listed in Appendix~\ref{app:current}.
Throughout the paper, all physical quantities, e.g. $\omega_r$ in \req{wind_rotor}, are in per unit unless otherwise stated. 



The DFIG is modeled through the dynamics of its
stator and rotor currents, expressed in a rotating d-q reference frame
as \cite{ugalde2013state} 
\begin{equation}\label{wind_current}
 \left\{
\begin{array}{rcl}
 \dot{i} &\hspace{-2.5mm}=&\hspace{-2.5mm} A_{i}(\omega_g)i + R_{i}|{\bf v}_{1}| + B_{i}(w+u) \vspace{2mm}\\
 \tau(i) &\hspace{-2.5mm}=&\hspace{-2.5mm} \displaystyle
  L_m \left(i_{qs}i_{dr} - i_{ds}i_{qr}\right)\vspace{1mm}\\
 {\bf e}_{1} &\hspace{-2.5mm}=&\hspace{-2.5mm}
  \gamma\left(|{\bf v}_{1}|i_{qs} + j |{\bf v}_{1}|i_{ds}\right)
\end{array}
\right., \quad
 i := 
\left[
\begin{array}{c}
 i_{dr}\\ i_{qr}\\ i_{ds}\\ i_{qs}\\
\end{array}
\right]
\end{equation}
where 
$i_{dr}$ and $i_{qr}$ are the d- and q-axis rotor currents, 
$i_{ds}$ and $i_{qs}$ are the d- and q-axis stator currents,
${\bf v}_1$ is phasor d- and q-axis stator voltage, 
${\bf e}_1 = p_1 + j q_1$ is the total effective wind power with $p_1$ being the active power and $q_1$ being the reactive power, 
$w \in \mathbb R^2$ is control input given by an internal controller (described below) to stabilize the wind generator, and 
$u \in \mathbb R^2$ is an additional control input whose design will be described in Section~\ref{sec:retrofit}. 
Each element of the term $w+u \in \mathbb R^2$ in \req{wind_current} represents d- and q-axis rotor voltage, respectively.

In this wind power plant model, the positive constant $\gamma$ in
\req{wind_current} represents the level of wind power penetration. 
In other words, a larger value of $\gamma$ indicates that a larger amount of wind power is injected to the grid. Later in our simulations we will numerically inspect the influence of this penetration level on the transient stability and dynamic performance of the Kundur system. 
Typically, both the d- and q-axis currents of the rotor need to be regulated to pre-computed setpoints based on a maximum power-point tracking algorithm for the operating wind speed. This regulation is needed  both for dynamic stability of the DFIG and for satisfying its steady-state power demand, and is achieved by the PI control of the currents as
\begin{subequations}\label{wind_pss}
\begin{equation}
\left\{
\begin{array}{l}
\begin{array}{rcl}
\dot{\xi}_{d} &\hspace{-2mm}=&\hspace{-2mm} \kappa_I(i_{dr} - i_{dr}^{\star})\\
w_{d} &\hspace{-2mm}=&\hspace{-2mm} \kappa_P(i_{d r} - i_{d r}^{\star})
 + \xi_{d} + z_{d}(\omega_g, i_{qs}, i_{qr})\\
\end{array}\vspace{1mm}\\
 z_{d}(\omega_g, i_{qs}, i_{qr}) := -(\omega_g - \omega_g^{\star})(L_m i_{qs} + L_r i_{qr})\\
\end{array}
\right.,
\end{equation}
and
\begin{equation}
\left\{
\begin{array}{l}
\begin{array}{rcl}
\dot{\xi}_{q} &\hspace{-2mm}=&\hspace{-2mm} \kappa_I(i_{qr} - i_{qr}^{\star})\\
w_{q} &\hspace{-2mm}=&\hspace{-2mm} \kappa_P(i_{q r} - i_{q r}^{\star})
 + \xi_{q} + z_{q}(\omega_g, i_{ds}, i_{dr})\\
\end{array}\vspace{1mm}\\
  z_{q}(\omega_g, i_{ds}, i_{dr}) := (\omega_g - \omega_g^{\star})(L_m i_{ds} + L_r i_{dr})\\
\end{array}
\right.
\end{equation}
\end{subequations}
where $w = [w_d, w_q]^{\sf T}$ is the controller output in \req{wind_current}, and $\xi_d\in\mathbb R$ and $\xi_q\in\mathbb R$ are the controller state, 
$\omega_g^{\star}$, $i_{dr}^{\star}$, $i_{qr}^{\star}$, are 
reference values of $\omega_g$, $i_{dr}$ and $i_{qr}$ achieving a given desirable steady-state power flow, 
$\kappa_P$ and $\kappa_I$ are the proportional and integral (PI) gains, $L_m$ and $L_r$ are the
magnetizing inductance and the rotor inductance of DFIG. 
The signals $z_d$ and $z_q$ are used to cancel out cross-coupling between d- and q-axis rotor voltages depending on the difference between the generator speed and synchronous speed;
for further details of this model, see \cite{anaya2014offshore}. 

\subsubsection{Synchronous Generators}\label{sec:sync}

 For $k\in\{2,\ldots,N\}$, the dynamics of the $k$-th synchronous generator model \cite{souvik2016time} consists of the electro-mechanical swing dynamics
\begin{equation}\label{model_sync1}
\hspace{0mm}\left\{\hspace{0mm}
 \begin{array}{rcl}
  \dot{\delta}_k &\hspace{-2mm}=&\hspace{-2mm} \bar{\omega} \omega_k\\
  m_k \dot{\omega}_k &\hspace{-2mm}=&\hspace{-2mm} p_{mk} - d_k \omega_k - |{\bf v}_k|\rho_k \sin(\delta_k - \angle {\bf v}_k) / \chi_k
 \end{array}
\right.
\end{equation}
and the electro-magnetic excitation dynamics
\begin{equation}\label{model_sync2}
\hspace{0mm}\left\{\hspace{0mm}
 \begin{array}{rcl}
  \tau_k \dot{\rho}_k&\hspace{-2mm}=&\hspace{-2mm} -\alpha_k \rho_k + \beta_k |{\bf v}_k|\cos(\delta_k - \angle {\bf v}_k) + \nu_k \\
  \tau'_k \dot{\nu}_k&\hspace{-2mm}=&\hspace{-2mm} -\nu_k - \kappa_k(|{\bf v}_k| -\mu_k)
 \end{array}
\right.
\end{equation}
where $\delta_k$ [rad] is the rotor angle relative to the coordinate system rotating at constant synchronous speed $\bar{\omega}$ [rad/s], 
$\omega_k$ is the rotor angular velocity relative to $\bar{\omega}$, 
$\rho_k$ is the internal voltage of the rotor, 
$\nu_k$ is the voltage of the excitation winding, 
$p_{mk}$ is the mechanical input, and 
${\bf v}_k$ is the complex terminal bus voltage, which is determined by the power flow between generation units and loads. 
For simplicity, we assume that $p_{mk}$ is a given constant value
achieving a desirable steady-state power generation for every synchronous machine $k \in \{2,\ldots, N\}$.
The constant parameters in \req{model_sync1}-\req{model_sync2} are as follows: $m_k$ and $d_k$ are inertia and damping coefficients of the rotor, 
$\tau_k$ is the open circuit time constant, $\tau'_k$ and $\kappa_k$ are the regulator gain and time constant,
$\chi_k$ is the d-axis sub-transient reactance, $\alpha_k$ and $\beta_k$ are the parameters depending on $\chi_k$ and the d-axis transient reactance. 
The total effective power generated by the $k$-th synchronous generator, consisting of both active and reactive powers, can be written as 
\begin{equation}\label{model_sync3}
 \begin{array}{rcl}
  {\bf e}_k &\hspace{-2mm}=&\hspace{-2mm} \rho_k|{\bf v}_k|(\sin \delta_k - \cos \angle {\bf v}_k) / \chi_k \vspace{1mm}\\
  &&\hspace{-2mm} + j\left( \rho_k^2 - \rho_k|{\bf v}_k|(\cos \delta_k - \sin \angle {\bf v}_k)\right)/\chi_k.
 \end{array}
\end{equation}

Typically generators are equipped with Automatic Voltage Regulators (AVR) that regulate their bus voltages  to setpoint values, as well as Power System Stabilizers (PSS) that ensure small-signal stability. A typical PSS can be represented as a speed-feedback controller 
\begin{equation}\label{model_pss}
 \mu_k = \mathcal C_k(\omega_k), \quad k \in \{2,\ldots,N\}
\end{equation}
where $\mathcal C_k(\cdot)$ is a dynamical map to be designed. 
In this paper, we use two stages of lead-lag compensators with one
highpass washout filter as shown in \cite{kundur1994power} to design $\mathcal C_k(\cdot)$. 

\subsubsection{Coupling between grid and wind plant}\label{sec:interconnection}
The dynamics of the synchronous generators and of the wind power plant are coupled to each other through power flow balance between the generators and the loads. This balance can be expressed as 
\begin{equation}\label{pow_balance}
 {\bf e} = \left({\bf Y}{\bf v}\right)^* \times {\bf v}
\end{equation}
where ${\bf e} \in \mathbb C^N$ and ${\bf v} \in \mathbb C^N$ are the stacked representations of ${\bf e}_k$ and ${\bf v}_k$, 
${\bf Y} \in \mathbb C^{N\times N}$ is the Kron-reduced network admittance matrix,
which involves load admittances, and the symbol $\times$ denotes element-wise multiplication. 
In the following, we refer to the power system outside the wind power plant as the 
{\it pre-existing grid model}. 
This nomenclature clearly acknowledges that in our analysis we treat the
grid with the synchronous generators as a nominal system, while  the
wind power plant is looked upon as a new addition to it. The goal is to
design a local controller for the wind power plant that can guarantee
stability and performance of the overall power system despite the intrusion of this new component.


\subsection{Motivating Example}\label{sec:motiv}
We next consider a motivating example for transient stability of the wind-integrated Kundur system modeled by \req{wind_rotor}-\req{pow_balance}. We assume 
$u=0$ in \req{wind_current}, $\kappa_P = 1$ and $\kappa_I = 5.2$ in \req{wind_pss}, and 
design a linear PSS $\mathcal C_k(\cdot)$ in \req{model_pss} such that the nonlinear model of the overall power system is stable.

\begin{figure}[t]
  \begin{center}
    \includegraphics[clip,width=80mm]{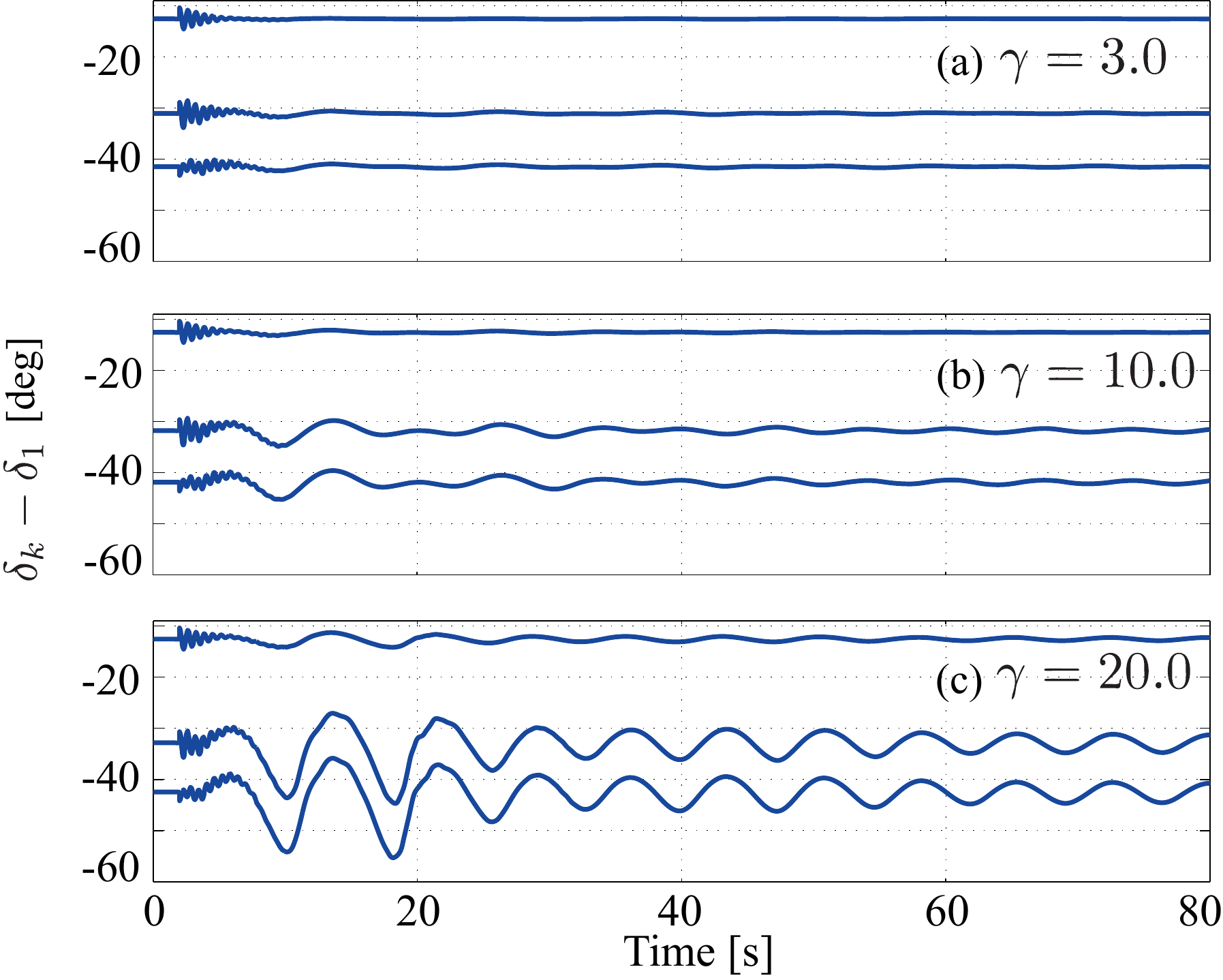}
    \caption{Comparison of $\delta_k-\delta_1$ in the case of $\gamma=3$, $10$, and $20$}
    \label{fig:wind}
  \end{center}
\vspace{-5mm}
\end{figure}

\begin{figure}[t]
  \begin{center}
    \includegraphics[clip,width=85mm]{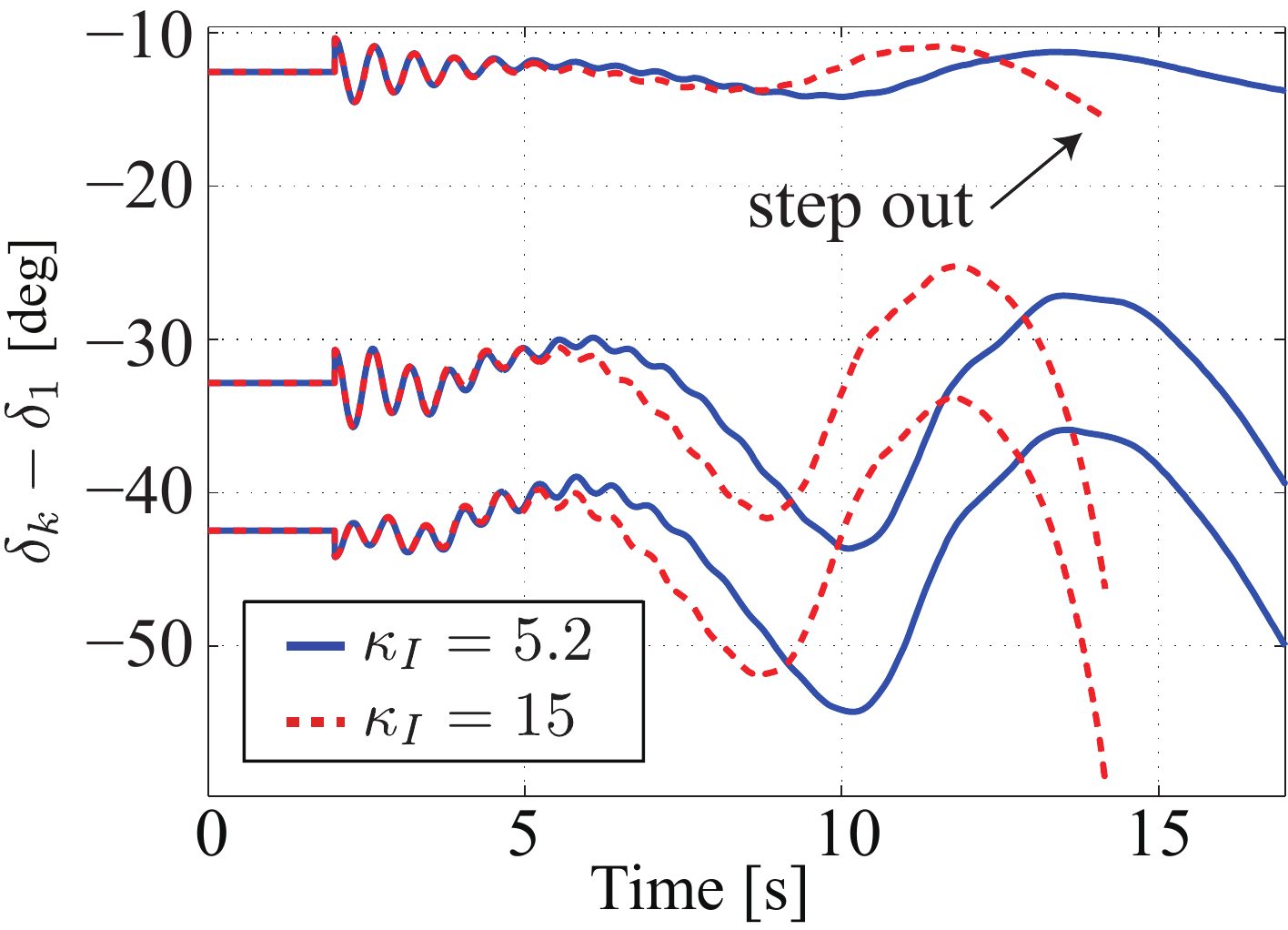}
    \caption{$\delta_k-\delta_1$ obtained by using the controller with $\kappa_I = 5.2$ and  $\kappa_I = 15$ in \req{wind_pss}}
    \label{fig:badpi}
  \end{center}
\vspace{-5mm}
\end{figure}

We suppose a situation where the system stays at an equilibrium during $t \in [0,2)$, and
subsequently, a line fault happens at $t=2$. 
First, we suppose a situation where $\gamma=3$ in \req{wind_current},
i.e., only a very small amount of wind power is injected to the grid. We
plot the synchronous generator angles relative to that of the first
synchronous generator, i.e., $\delta_k-\delta_1$, in
Fig.~\ref{fig:wind}(a). We can see from this figure that the
trajectories tend to go to steady-state values within 10 [sec].

However, when $\gamma$ increases to 10 or 20, we can see from Figs.~\ref{fig:wind}(a)-(c) that the behavior of the phase angle differences becomes more oscillatory as the penetration level gets higher. 
We also linearize the system model \req{wind_rotor}-\req{pow_balance} for different values of $\gamma$ in the range $[3, 20]$. 
Then, we found that two eigenvalues move from $-0.055\pm 1j$ to $-0.014\pm 0.8j$ as the value of $\gamma$ increases. In addition, these eigenvalues go across the imaginary axis if $\gamma$ is larger than $25.8$. 
This result shows how wind penetration beyond a certain limit can induce transient instability in the Kundur model. 

One potential way to combat this would be to increase the gains of the PI controllers in \req{wind_pss}.
However, such a high gain controller may in fact end up destabilizing the whole system by stimulating negative coupling effect coming from the pre-existing grid model. 
Indeed, for $\kappa_I = 15$, the system becomes unstable as shown in
Fig.~\ref{fig:badpi}. 
Therefore, to mitigate the adverse impacts of wind power integration to the power grid, we need a systematic control mechanism by which
the behavior of the wind power plant can be regulated in a desired way without causing any grid instability.
We next propose such a systematic control design for the wind power plant using only its local output feedback.

\section{Retrofit Control}\label{sec:problem}
\subsection{Retrofit Control of Wind-Integrated Power System}\label{sec:retrofit}
Towards improvement of transient stability of wind-integrated
power systems, we aim at designing a {\it retrofit controller}
\cite{sadamoto2016retrofitting}, which sends additional control input $u$ in
\req{wind_current}. 
We will show that this controller not only retains stability but also enhances the damping of the line power flows and generator frequencies in transience.  
In the following, we assume that $|{\bf v}_1|$ and $\omega_r$ in
\req{wind_rotor} are measurable in addition to $\omega_g$ and $i$ used
by the internal controller \req{wind_pss}. 
Summarizing \req{wind_rotor}-\req{wind_pss}, we can rewrite the wind power plant dynamics as
\begin{equation}\label{sigma_wind}
 \Sigma:~ 
\left\{
 \begin{array}{rcl}
  \dot{x}&\hspace{-2mm}=&\hspace{-2mm} F(x) + R|{\bf v}_1| + Bu\\
  {\bf e}_1&\hspace{-2mm}=&\hspace{-2mm} g(x, |{\bf v}_1|) \\
 \end{array}
 \right. 
\end{equation}
where 
\[
 x := \left[
\begin{array}{c}
 \eta \\ i \\ \xi_d \\ \xi_q
\end{array}
\right]\in\mathbb R^9, \quad
 R := \left[
\begin{array}{c}
 0 \\ R_i \\ 0 \\ 0
\end{array}
\right], \quad
 B := \left[
\begin{array}{c}
 0 \\ B_i \\ 0 \\ 0
\end{array}
\right].
\]
It should be noted that these nonlinear dynamics depend on 
\begin{equation}\label{defy}
 y := [\omega_r, \omega_g, i^{\sf T}]^{\sf T} = Cx
\end{equation}
where $C$ is compatible with $y$. 
Thus, the nonlinear function $F(x)$ in \req{sigma_wind} has a form of
\begin{equation}\label{jacob}
 F(x) = Ax + f(Cx)
\end{equation}
where $A$ is the Jacobian matrix of $F(x)$ at an operating point $x^{\star}$, i.e., 
\begin{equation}\label{approxA}
A := \frac{\partial F}{\partial x}(x^{\star})
\end{equation}
and $f(\cdot)$ is the residue defined as $f(Cx) := F(x) - Ax$. 
We consider that  $x^{\star}$ is given as an equilibrium of $x$ in \req{sigma_wind} such that
\begin{equation}\label{equil}
 0 = F(x^{\star}) + R|{\bf v}_1^{\star}|, \quad 
  {\bf e}_1^{\star} = g(x^{\star}, |{\bf v}_1^{\star}|)
\end{equation}
for the desirable bus voltage ${\bf v}_1^{\star}$ and power ${\bf e}_1^{\star}$. 
In addition, we denote $y^{\star} := Cx^{\star}$. 
In this setting, a retrofit controller for the wind-integrated power
system is given as follows \cite{sadamoto2016retrofitting}:



\begin{proposition}\vspace{3pt}\label{lemfun}
Assume that the interconnection of the pre-existing grid model in
 \req{model_sync1}-\req{pow_balance} and wind power plant $\Sigma$ in
 \req{sigma_wind}-\req{approxA} is stable. 
Consider a dynamical map $\mathcal K(\cdot)$ stabilizing 
\begin{equation}\label{dessta}
\dot{\zeta}=A\zeta+B\mathcal K(C\zeta).
\end{equation}
Then, the feedback system of \req{model_sync1}-\req{approxA}  interconnected by the control input
\begin{equation}\label{fedcon}
u=\mathcal K(y-\hat{y})
\end{equation}
where $\hat{y}$ is given by the dynamic compensator:
\begin{equation}\label{syscon}
\hat{\Sigma}:
\left\{
\begin{array}{ccl}
\dot{\hat{x}}&\hspace{-6pt}=&\hspace{-6pt}A\hat{x}+f(y - y^{\star})+R(|{\bf v}_1|-|{\bf v}_1^{\star}|)\\
\hat{y}&\hspace{-6pt}=&\hspace{-6pt}C\hat{x} + y^{\star}
\end{array}
\right.
\end{equation}
is stable.
\end{proposition}\vspace{3pt}

The combination of the controller $\mathcal K(\cdot)$ and the compensator
$\hat{\Sigma}$ in Proposition~\ref{lemfun} is referred to as a {\it retrofit controller}. 
This controller is implemented in the current control loop of the DFIG exactly in a parallel connection to the pre-existing PI controllers. 
The retrofit controller needs the bus voltage magnitude $|{\bf v}_1|$,
and only a part of the wind power plant state for feedback, namely the angular velocity of rotor and generator $\omega_r$, $\omega_g$ and the
DFIG currents $i$.

Following the design philosophy provided in \cite{sadamoto2016retrofitting}, the controller $\mathcal K(\cdot)$ can be tuned for improving dynamic performance of the wind-integrated system.
For simplicity, we suppose that $\mathcal K(\cdot)$ is a linear map.
We can show that there exists a class-$\mathcal K$ function $\beta(\cdot)$ satisfying 
\begin{equation}\label{performance}
 \|x\|_{\mathcal L_2} \leq \beta(\|\zeta\|_{\mathcal L_2})
\end{equation}
where $\zeta$ obeys \req{dessta} with $\zeta(0) = x(0)$. 
Thus, if a map $\mathcal K(\cdot)$ is tuned such that
$\|\zeta\|_{\mathcal L_2}$ is made smaller by any conventional optimal control design such as linear quadratic regulator (LQR), 
then the performance in terms of $\|x\|_{\mathcal L_2}$, which
would evaluate damping performance of the wind power plant, can
also be made smaller in the sense of the bound in \req{performance}. 
This theoretical guarantee for the performance improvement is what distinguishes our proposed retrofit control from the current state-of-art wind power control methods, most of which are based on {\it adhoc} tuning of PID controllers.  Later in our simulations we will show that this additional controller is capable of improving the damping characteristics of the pre-exisiting grid model as well. 

One may wonder if the control $u = \mathcal K(y)$, i.e., the control without using the compensator $\hat{\Sigma}$, can improve the performance of the wind power plant. 
However, in general, such a control does not have any guarantee of performance improvement, and to make matters worse, it may 
pose serious threat to the power system stability despite sufficient tuning of the PSS and PI controllers.
This is because the neglected dynamics of the wind-integrated system model, which is not only the nonlinearity $f(\cdot)$ in \req{approxA}, but also the dynamics of the pre-existing grid, are stimulated by the control $u = \mathcal K(y)$.
The compensator $\hat{\Sigma}$ in Proposition~\ref{lemfun} sends a compensation signal $\hat{y}$ in \req{fedcon} to prevent this stimulation. 
Hence, owing to this compensation we can theoretically guarantee the stability of the whole closed-loop system.

Another important point to note here is that we can design 
the controller $\mathcal K(\cdot)$ and the compensator $\hat{\Sigma}$ without
knowing the pre-existing grid model \req{model_sync1}-\req{pow_balance}. 
Thus, this retrofit controller design is extremely convenient to implement in practical large-scale power systems where it is almost impossible to have accurate knowledge of the exact model of the entire grid.

\begin{figure}[t]
\begin{center}
\includegraphics[width=85mm]{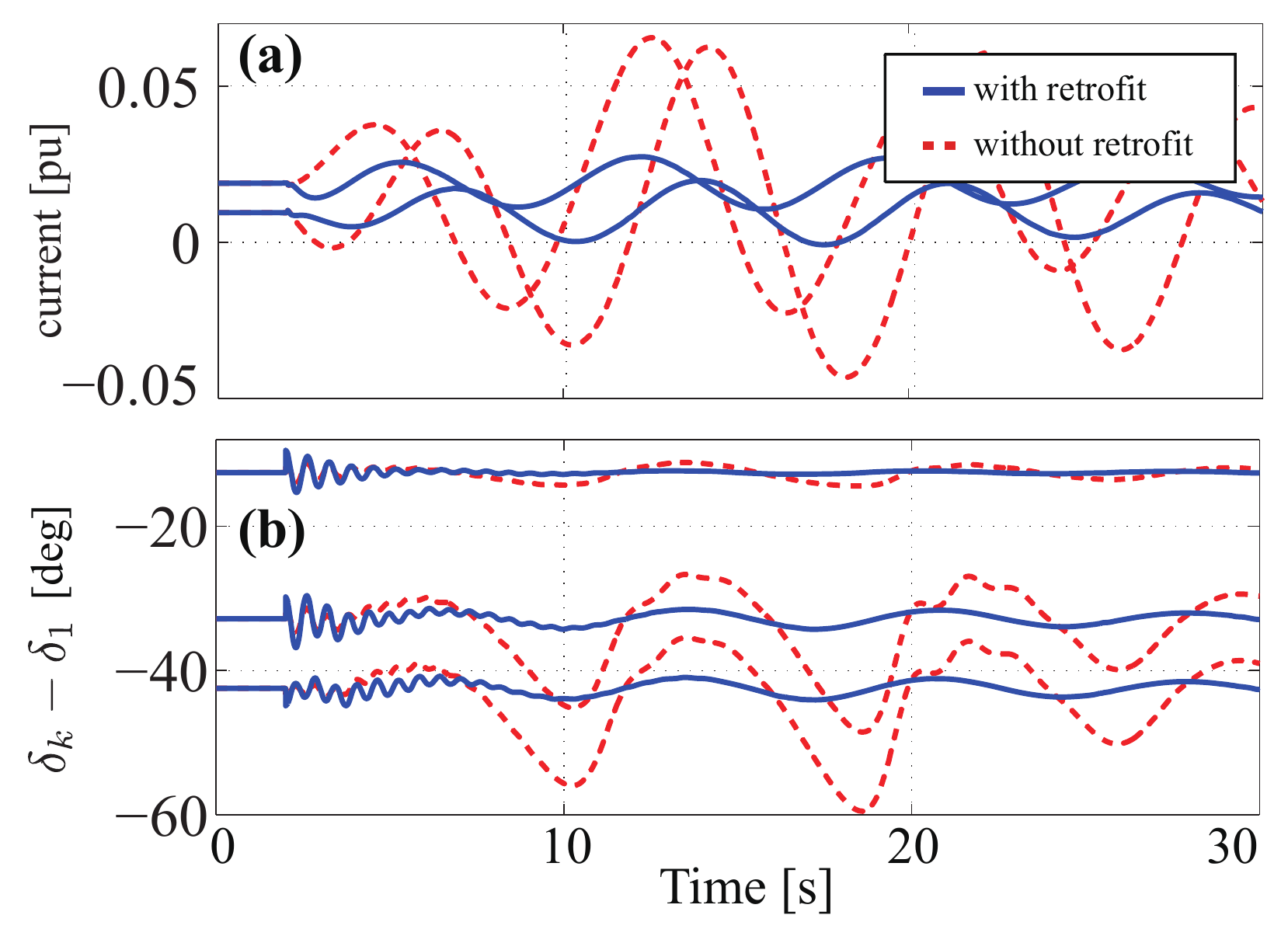}
\end{center}
\caption{Upper subfigure: $i_{ds}$ and $i_{qs}$. Lower subfigure: Synchronous generator angles relative to that of the first synchronous generator}
\vspace{-5mm}
\label{fig:sim02}
\end{figure}

\begin{figure*}[!t]
\vspace{-5mm}
\begin{center}
\includegraphics[width=176mm]{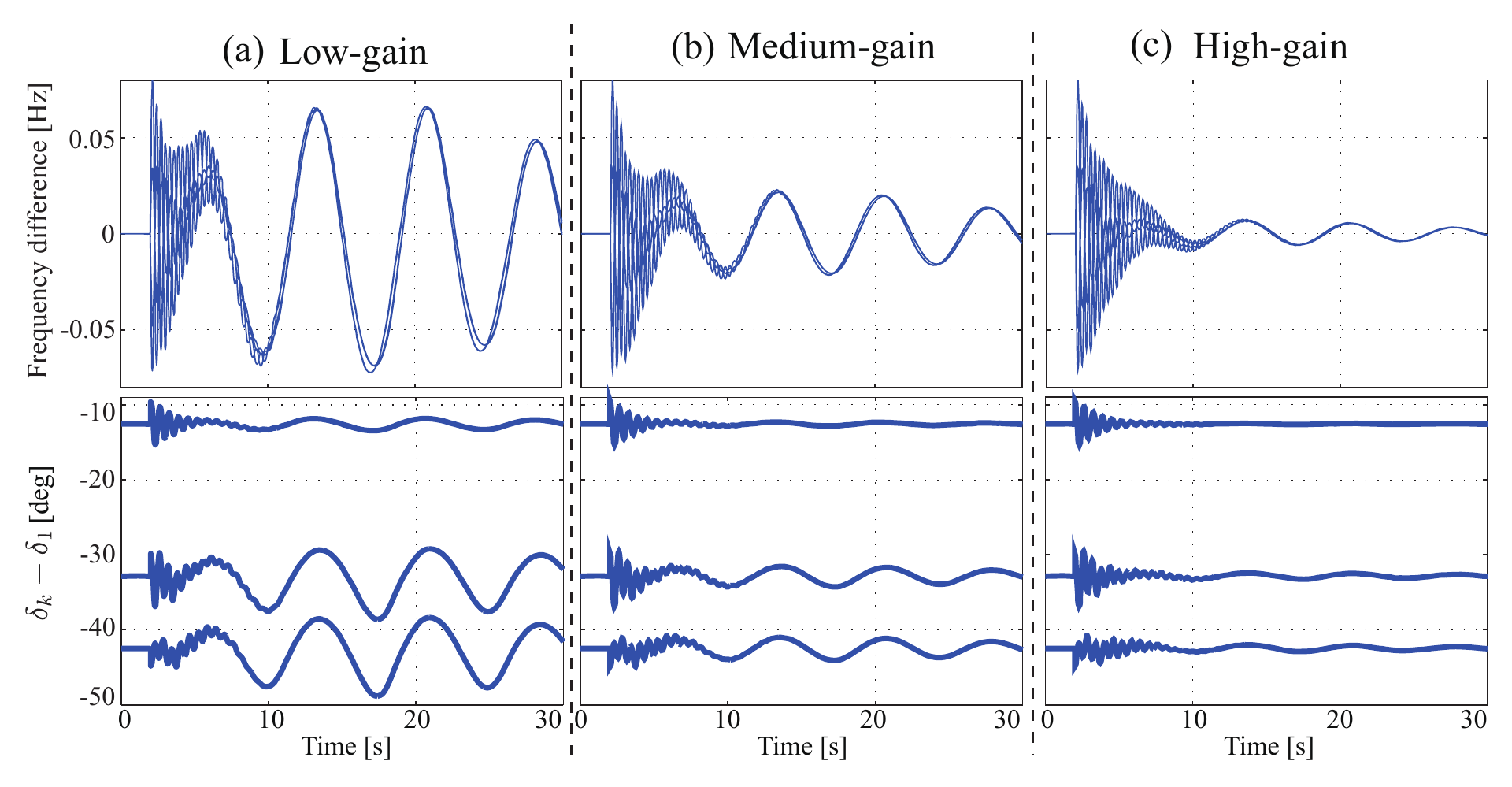}
\end{center}
\vspace{-5mm}
\caption{Frequency differences and angles relative to the first
 synchronous generator angle. }
\vspace{-2mm}
\label{fig:sim03}
\end{figure*}

\begin{figure*}[!t]
\normalsize
\setcounter{MYtempeqncnt}{\value{equation}}
\setcounter{equation}{17}
\begin{equation}\label{ABCDelectrical}
 A_{i}(\omega_g) := \left[
\begin{array}{cccc}
 -\frac{R_rL_s}{K_1}& 1-\omega_g\frac{K_4}{K_1}& \frac{R_sL_m}{K_1} &-\omega_g\frac{K_3}{K_1}\\
 -1+\omega_g \frac{K_4}{K_1} & -\frac{R_rL_s}{K_1} & \omega_g \frac{K_3}{K_1} & \frac{R_sL_m}{K_1} \\
 \frac{R_rL_m}{K_1} & \omega_g\frac{K_5}{K_1} & -\frac{R_sL_r}{K_1} & 1+\omega_g\frac{K_2}{K_1}\\
 -\omega_g\frac{K_5}{K_1} & \frac{R_rL_m}{K_1} & -1-\omega_g\frac{K_2}{K_1} &  -\frac{R_sL_r}{K_1} 
\end{array}
\right], \quad 
 R_{i} := \left[
\begin{array}{c}
0 \\ \frac{L_m}{K_1} \\ 0 \\ -\frac{L_r}{K_1}
\end{array}
\right] ,\quad B_{i} := \left[
\begin{array}{cc}
-\frac{L_s}{K_1} & 0\\
0 & -\frac{L_s}{K_1} \\
\frac{L_m}{K_1} & 0\\
0 &\frac{L_m}{K_1} \\
\end{array}
\right]
\end{equation}
\vspace{-5mm}
\end{figure*}
\setcounter{MYtempeqncnt}{\value{equation}}
\setcounter{equation}{17}

\section{Numerical Simulation}\label{sec:numerc}

In this section we validate our proposed control design by applying it to the wind-integrated Kundur power system model of Fig.~\ref{fig:example}, modeled by equations \req{wind_rotor}-\req{pow_balance}. We choose the wind injection limit $\gamma = 20$ in \req{wind_current}.
As shown in Section~\ref{sec:motiv}, a larger amount of wind penetration tends to induce significant oscillatory behavior. 
To mitigate this adverse impact, we construct a retrofit
controller $\{\mathcal K(\cdot), \hat{\Sigma}\}$, where $\hat{\Sigma}$
is given as \req{syscon} and 
$\mathcal K(\cdot)$ in \req{fedcon} is designed as an observer-based LQR.
In what follows, we consider the same fault at time $t=2$ as in Section~\ref{sec:motiv}. 

In Fig.~\ref{fig:sim02}(a), the blue solid lines depict the d- and q-axis stator currents in presence of the retrofit controller while the red dotted lines depict those without the controller. 
The reason why the blue response has an oscillation mode having a frequency around $0.1$[Hz] is that this mode is uncontrollable by the control input $u$ in \req{sigma_wind}. 
By comparing the blue lines with the red lines, 
we can see that the stator current fluctuation can be suppressed by the retrofit controller. 
The other state variables of the wind power plant,  e.g., the rotor current, generator angular velocity, show a similar improved transient response. 
We also compare the closed-loop response of the entire power system model with the wind plant with and without the retrofit controller.
In Fig.~\ref{fig:sim02}(b), the blue solid lines depict the synchronous generator angles relative to that of the first generator, i.e., $\delta_k - \delta_1$, in presence of the retrofit controller, while the red dotted lines also depict the case without the controller. 
This subfigure implies that the magnitude of the oscillation of the relative angles can be suppressed by the retrofit controller. 

To investigate the variation of the damping performance of the closed-loop system, 
we design several $\mathcal K(\cdot)$ by varying the weight for the
LQR. 
Figs.~\ref{fig:sim03}(a)-(c) show the frequency difference and $\delta_k
- \delta_1$ corresponding to the case with low, middle, and high gain
controllers. 
As we can see from these subfigures, the damping performance of the closed-loop system monotonically improves as we apply higher gains in the controller $\mathcal K(\cdot)$.

\section{Conclusions}\label{sec:concl}
In this paper, we have proposed retrofit control of wind-integrated
power systems, where the control action is actuated based on local feedback
of the wind power plant state only. 
As the level of wind penetration increases, the transient response of
line flows and generator frequencies after a fault may become highly
oscillatory. Tuning PI controllers in the current control loops of the
DFIG may be able to control these oscillations to some extent, but
increasing these gains eventually tend to destabilize the entire power system. Our proposed retrofit controller circumvents this difficulty, and retains both stability and dynamic performance despite very high levels of wind penetration. 
Furthermore, the controller design can be performed without
explicit knowledge of the dynamics of the system other than the wind
power plant. 
We have shown the efficiency of the proposed control through a Kundur 4-machine power system model with a wind power plant at the
intermediate bus. 

\section{Acknowledgement}
This research was supported by CREST, JST.
The work of the second author was partly supported by NSF ECCS grants 123084 and 1054394.
The authors are deeply grateful to Sayak Mukherjee from North Carolina State University for his help with the numerical simulations. 

\appendix

\subsection{Wind Power Plant Parameters}\label{app:current}
The parameters of the wind turbine in \req{wind_rotor} are summarized as follows:
$m_r = 103$ [${\rm s}^2$],
$m_g = 7.3 \times 10^-4$ [${\rm s}^2$],
$d_r = 0.05$ [rad/s],
$d_g = 5.7\times 10^{-6}$ [rad/s],
$k_c = 0.27$ [s],
$d_c = 1.78$ [rad/s], 
$p_a = 2.5\times 10^{-2}$ , and
$n_g= 90$.
Note that all the values are rated at the system capacity 100 [MVA].

The matrices of DFIG in \req{wind_current} of the wind power plant rated at $1.8$MW and $575$V are defined in \req{ABCDelectrical} where
$L_s = 4.8365$ and $L_r = 4.8344$ are the stator and rotor inductance, 
$R_s = 0.0111$ and $R_r = 0.0108$ are the stator and rotor resistance, $L_m = 4.6978$ is the magnetizing inductance, 
$K_1 := (L_sL_r - L_m^2)$, $K_2 := L_m^2$,  $K_3 := L_sL_m$, $K_4 :=
L_rL_s$, $K_5 := L_rL_m$, and $n_p = 4$ is the number of electrical poles of the DFIG.

\bibliographystyle{IEEEtran}
\bibliography{reference}
\end{document}